\documentclass[fleqn,twoside]{article}
\usepackage{espcrc2}

\usepackage{graphicx}
\usepackage[figuresright]{rotating}

\newcommand{\AmS}{{\protect\the\textfont2
  A\kern-.1667em\lower.5ex\hbox{M}\kern-.125emS}}

\title{\begin{flushright}
\bf \small{ DFUB 10/2002 \\
\bf 24/9/02}\\
\end{flushright}
Search for GUT Magnetic Monopoles with the MACRO Experiment at the Gran Sasso Lab.}

\author{Giorgio Giacomelli, for the MACRO Collaboration \thanks{See ref. 1 for the list of MACRO collaborators and Institutions} \address{Dipartimento di Fisica dell'Universit\`a di Bologna and INFN, I-40127 Bologna, Italy\\	
e-mail: giacomelli@bo.infn.it \\
\vspace{0.3cm}
\bf Contribution to the ICHEP 2002 Conference, Amsterdam, 24-31 July 2002 }}

\begin{document}

\begin{abstract}
The final results of the MACRO experiment on the search for GUT supermassive magnetic monopoles in the 
penetrating cosmic radiation are presented and discussed. The 90\% CL upper limits are at the level of 
1.5 - \(2 \times 10^{-16}\) \(cm^{-2}\) \(s^{-1}\) \(sr^{-1}\) for \(0.0004 < \beta < 1\). Similar limits were obtained for nuclearites. Limits for the MM catalysis of proton decay are at \(3 - 4 \times 10^{-16}\) 
\(cm^{-2}\) \(s^{-1}\) \(sr^{-2}\) for \(10^{-4}< \beta < 6 \cdot 10^{-3}\).

\vspace{1pc}
\end{abstract}

% typeset front matter (including abstract)
\maketitle

%\section{FORMAT}
One of the primary aims of the MACRO detector at the Gran Sasso lab. (at an average depth of 3700 mwe) was the search in the penetrating cosmic radiation for the supermassive magnetic monopoles (MMs) predicted by Grand Unified Theories of the electroweak and strong interactions; the search was planned for a sensitivity 
level well below the 
Parker bound \((\sim 10^{-15} \; cm^{-2} \; s^{-1} \; sr^{-1})\) and for a large range of velocities, 
\(4 \cdot 10^{-5} < \beta < 1, \; \beta \)= v/c.

MACRO used three different types of detectors: liquid scintillators, limited steamer tubes and nuclear track detectors (NTDs; CR39 and Lexan) arranged in a 
closed-box modular structure, divided into a lower and an upper 
(\lq\lq Attico'') part. The overall dimensions of the apparatus were \(76.5 \times 12 \times 9.3 \; m^{3}\). MACRO took 
data from early 1989 till december 2000 [1]. 

The response of the three types of detectors to slow and fast particles was experimentally studied [2 - 4]. 
The use of three types of subdetectors with redundant electronics, in stand alone and combined modes, ensured redundancy of information, cross-checks and indipendent signatures for possible MM candidates; it also allowed to cover the complete \( \beta-range\), \( 4 \times 10^{-5} - 1\). 

The searches performed with the liquid scintillator subdetector used different specialized triggers 
to cover the \(10^{-4} < \beta < 4 \times 10^{-2}\) range [2]; a custom made 200 MHz wave form digitizer was used for the lower \(\beta-range\), while the 
PHRASE system was used in the intermediate \(\beta-range\).

The searches performed in the low \(\beta-region\) with the limited streamer tubes were based on the search for single tracks and on the measurement of the velocity by means of a \lq\lq time track'' [3].

The searches with the CR39 NTDs required chemical etching of each individual sheet 
\((24.5 \times 24.5 \times 0.14 \; cm^3)\), double scannings, measurements 
and eventual cross checks with other sheets of the same module. We analized 845.5 \(m^2\) of sheets, exposed for an average time of 9.5 y [4].

Combined analyses with 2 or 3 types of detectors were performed as described in ref. [5]. 

No MM candidate was found; the obtained limits 
were presented and discussed at many conferences and compared with other results [6] [7]. 

Fig. 1 
shows the 90\% C.L. flux upper limits for \(g = g_{D}\) MMs (one unit Dirac 
magnetic charge) obtained by indipendent methods.
The final limit, computed by combining the indipendent limits obtained by the 
different analyses, is shown in Figs. 1 and 2; 
the MACRO limit is compared with previous limits in Fig. 2 [8].

\begin{figure}[htb]
\vspace{-2cm}
\includegraphics[scale=0.35]{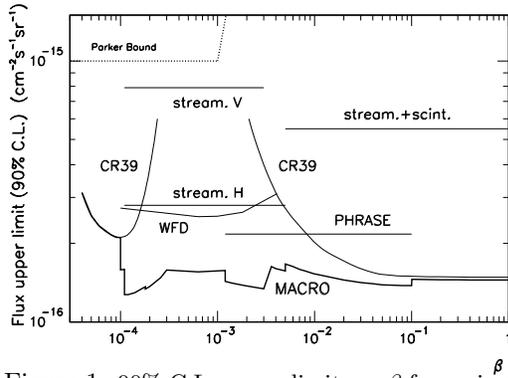}
%\framebox[55mm]{\rule[-21mm]{0mm}{43mm}}
\vspace{-1.2cm}
\caption{\small{90\% C.L. upper limits vs \(\beta\) for an isotropic flux of MMs obtained from different searches with the 3 subdetectors (both in stand alone and combined ways), and the MACRO global limit.}}
\label{fig:l}
\vspace{-0.5cm}
\end{figure}

\begin{figure}[htb]
\vspace{-2cm}
\includegraphics[scale=0.35]{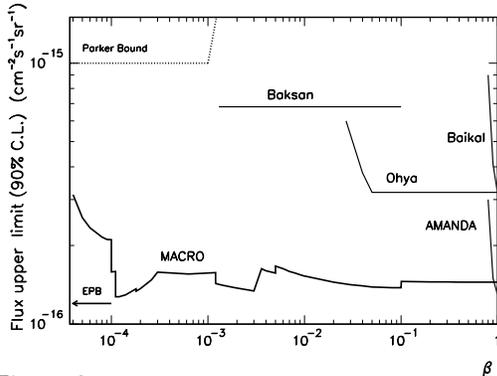}
%\framebox[55mm]{\rule[-21mm]{0mm}{43mm}}
\vspace{-1.2cm}
\caption{\small{The global MACRO MM upper limit compared with limits from other experiments [8].}}
\label{fig:2}
\vspace{-0.7cm}
\end{figure}

\begin{figure}[htb]
\includegraphics[scale=0.35]{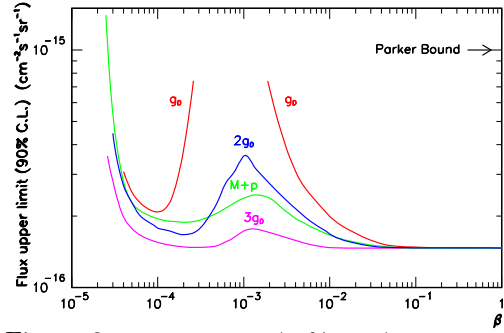}
%\framebox[55mm]{\rule[-21mm]{0mm}{43mm}}
\vspace{-1.2cm}
\caption{\small{Upper limits (90\% C.L.) for an isotropic flux of MMs obtained with the CR39 subdetector of MACRO, for poles with magnetic charge \(g=g_D,\; 2g_D, \;3g_D\) and for M+p composites.}}
\label{fig:3}
\vspace{-0.5cm}
\end{figure}

\begin{figure}[htb]
\vspace{-0.5cm}
\includegraphics[scale=0.35]{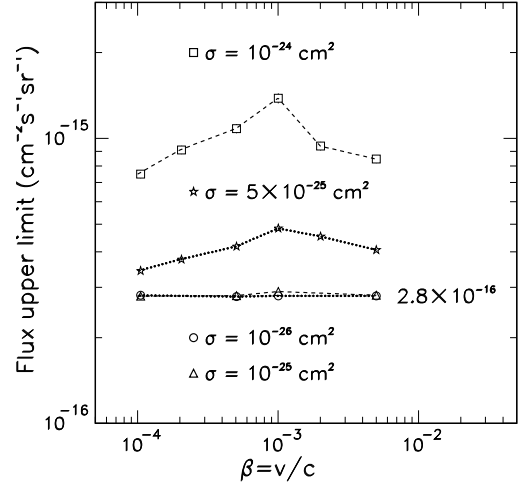}
%\framebox[55mm]{\rule[-21mm]{0mm}{43mm}}
\vspace{-1cm}
\caption{\small{Upper limits for a MM flux as a function of the MM velocity for various catalysis cross sections.}}
\label{fig:4}
\vspace{-0.6cm}
\end{figure}

\begin{figure}[htb]
\vspace{-1cm}
\scalebox{0.35}{\includegraphics{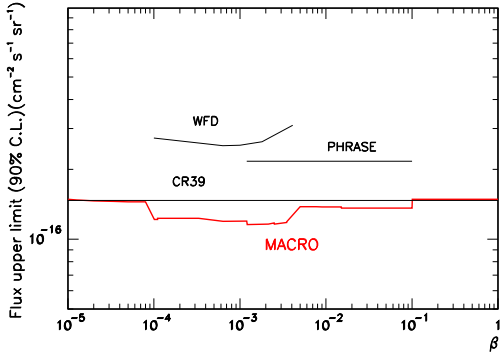}}
%\framebox[55mm]{\rule[-21mm]{0mm}{43mm}}
\vspace{-1cm}
\caption{\small{The 90\% C.L. upper limits for an isotropic flux of nuclearites obtained with the liquid scintillators (WFD and PHRASE) and the CR39 subdetectors. The bottom line is the MACRO global limit.}}
\label{fig:5}
\end{figure}

\begin{figure}[htb]
\vspace{-0.8cm}
\scalebox{0.35}{\includegraphics{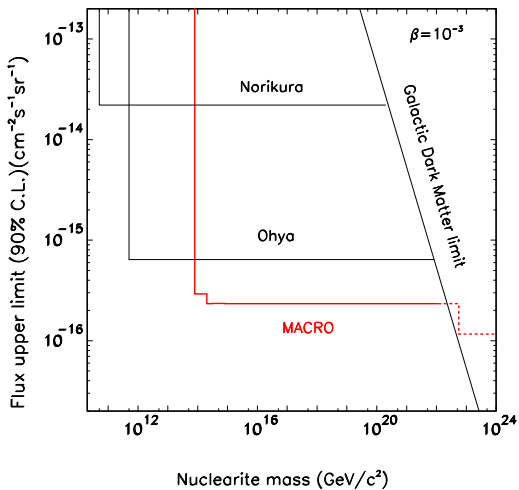}}
%\framebox[55mm]{\rule[-21mm]{0mm}{43mm}}
\vspace{-1cm}
\caption{\small{The MACRO flux upper limit for nuclearites with \(\beta=10^{-3}\) at ground level vs nuclearite mass, compared with those obtained by other experiments and with the dark matter bound. }}
\label{fig:6}
\vspace{-0.7cm}
\end{figure}

Fig. 3 shows the flux upper limits obtained with the CR39 nuclear track detector for MMs with different magnetic 
charges, \(g=g_{D}\), \( 2g_{D}\), \(3g_{D} \) and for M + p composites.

The interaction of the GUT monopole core with a nucleon can lead to a reaction in which the nucleon decays (monopole catalysis of nucleon decay), f. e. \( M + p \rightarrow M + e^+ + \pi^0\) [9]. The cross 
section for this process is of the order of the core size, \( \sigma \sim 10^{-56} cm^2\), 
practically negligible. But the catalysis process could proceed via the Rubakov-Callan mechanism with a cross section of the order of the strong interaction cross section [10]. We developed a dedicated analysis procedure aiming to detect nucleon decays induced by the passage of a GUT MM in our streamer tube system. The flux 
upper limit results of this search as a function of the MM velocity and of the catalysis cross section is shown in Fig. 4 [11].

MMs with intermediate masses, \(10^{5} \; < m_{M} < 10^{12}\) GeV, could be present in the cosmic radiation; underground and underwater experiments have a limited sensitivity to such MMs.

As a byproduct of our scintillator and nuclear track detector searches, we obtained upper limits for a flux of nuclearites and of Q-balls in the penetrating cosmic radiation.

Nuclearites (Strangelets, Strange Quark Matter) should consist of aggregates of u, d and s quarks in almost equal proportion. They could have been produced 
shortly after the Big Bang or in violent astrophysical processes, and may have survived as remnants; they could be part of the cold dark matter and would 
have typical galactic velocities, \( \beta \sim 10^{-3}\). They should yield a 
large amount of light in scintillators and large Restricted Energy Losses in the nuclear track detectors. Fig. 5 shows our upper limits for an isotropic flux of nuclearites obtained with the scintillator and the CR39 subdetectors; also the MACRO combined limit is shown [12]. Fig. 6 shows a compilation of limits for a flux of nuclearites compared with the dark matter limit, assuming a velocity at ground level of \(\beta = 10^{-3}\).

Q-balls should be aggregates of squarks \( \tilde{q}\), sleptons \( \tilde{l}\) and Higgs fields [13]. The scalar condensate inside a Q-ball core should have a global baryon number Q.
There could exist neutral and charged Q-balls and they could be possible cold 
dark matter candidates. Charged Q-balls could have been detected, with reduced sensitivity, by our scintillators and nuclear track detectors.

In conclusion, we have presented the final results of the MACRO searches for superheavy MMs in the penetrating cosmic radiation. The flux limits are at the level of \(1 - 2 \times 10^{-16} \; cm^{-2} \; s^{-1} \; sr^{-1}\) for 
\( \beta > 4 \times 10^{-5}\). The obtained limits are the best existing and cover the widest \( \beta-range\). It will be difficult to do much better since 
one would require detectors of much larger areas.

Similar considerations apply to massive nuclearites with \( \beta \sim 10^{-3}\) and to charged Q-balls.

I thank the colleagues of the MACRO collaboration, in particular those from Bologna. I thank ms. Anastasia Casoni for her cooperation.

%Text should be produced within the dimensions shown on these pages:
%each column 7.5 cm wide with 1 cm middle margin, total width of 16 cm
%and a maximum length of 19.5 cm on first pages and 21 cm on second and
%following pages. The \LaTeX{} document class uses the maximum stipulated
%length apart from the following two exceptions (i) \LaTeX{} does not
%begin a new section directly at the bottom of a page, but transfers the
%heading to the top of the next page; (ii) \LaTeX{} never (well, hardly
%ever) exceeds the length of the text area in order to complete a
%section of text or a paragraph. Here are some references:
%\cite{Scho70,Mazu84}.

%\subsection{PostScript figures}

%Instead of providing separate drawings or prints of the figures you
%may also use PostScript files which are included into your \LaTeX{}
%file and printed together with the text. Use one of the packages from
%\LaTeX's \texttt{graphics} directory: \texttt{graphics},
%\texttt{graphicx} or \texttt{epsfig}, with the \verb|\usepackage|
%command, and then use the appropriate commands
%(\verb|\includegraphics| or \verb|\epsfig|) to include your PostScript
%file.

%The simplest command is: \newline
%\verb|\includegraphics{file}|, which inserts the
%PostScript file \texttt{file} at its own size. The starred version of
%this command: \newline
%\verb|\includegraphics*{file}|, does the same, but clips
%the figure to its bounding box.

%With the \texttt{graphicx} package one may specify a series of options
%as a key--value list, e.g.:
%\begin{tabular}{@{}l}
%\verb|\includegraphics[width=15pc]{file}|\\
%\verb|\includegraphics[height=5pc]{file}|\\
%\verb|\includegraphics[scale=0.6]{file}|\\
%\verb|\includegraphics[angle=90,width=20pc]{file}|
%\end{tabular}


\begin{thebibliography}{9}
%\bibitem{Scho70} S. Scholes, Discuss. Faraday Soc. No. 50 (1970) 222.


\bibitem{ } S.Ahlen et al., Nucl. Instr. Meth. Phys. Res. A324(1993)337; M.Ambrosio et al., Nucl. Instr. Meth. Phys. Res. A486(2002)663.

\bibitem{ } S.Ahlen et al., Phys. Rev. Lett. 72(1994)608; M.Ambrosio et al., Astroparticle Phys. 6(1997)113.
\bibitem{} M.Ambrosio et al., Astroparticle Phys. 4(1995)33.
\bibitem{} S.Cecchini et al, Il Nuovo Cimento A109(1996)1119; Radiat. Meas. 34(2001)54; G.Giacomelli et al., Nucl. Instr. Meth. A411(1998)41.
\bibitem{} M.Ambrosio et al., hep-ex/0110083.
\bibitem{} M.Ambrosio et al., Phys. Lett. B406(1997)249; hep-ex/0206027.
\bibitem{} G.Giacomelli et al., hep-ex/0002032; hep-ex/0005041; hep-ex/0112009.
\bibitem{} E.N.Alexeyev et al. (\lq\lq Baksan''), \(21^{st}\) ICRC, 
10(1990)83. S.Orito et al., (\lq\lq Ohya''), Phys. Rev. Lett. 66(1991)1951. 
	V.A.Balkanov et al., (\lq\lq Baikal'',  \lq\lq Amanda'' ), 
	Nucl. Phys. B91(2001)438.
\bibitem{} G.Giacomelli et al., hep-ex/0004019. 
\bibitem{} V.A.Rubakov, JETP Lett. B219(1981)644. G.G.Callan, Phys. Rev. 
	D26(1982)2058.
\bibitem{} M.Ambrosio et al., hep-ex/0207024, Eur. Phys. J. C26(2002)163.
\bibitem{} S.Ahlen et al., Phys. Rev. Lett. 69(1992)1860; M.Ambrosio et al., 
	Eur. Phys. J. C13(2000)453.
\bibitem{} S.Coleman, Nucl. Phys. B262(1985)293. A.Kusenko, Phys. Lett. B405(1997)108.
\end{thebibliography}
\end{document}